# Nano-robotic based Thrombolysis: Dissolving Blood Clots using Nanobots

Puru Malhotra
*Department of Information Technology*
*Maharaja Agrasen Institute of Technology*
New Delhi, India
puru2991999@gmail.com

Nimesh Shahdadpuri
*Department of Electronics and Communication*
*Netaji Subhas University of Technology*
New Delhi, India
shahdadpuri.nimesh@gmail.com

*Abstract*— *Medicine has taken a big jump in its advancement in the past few years. With the introduction of technology and the synthesis of personalized medicines, doctors have been successful in treating even the deadliest of the diseases. Though surgeries, which have been flourishing with a good success rate in the past, have become feasible and they are being regularly performed to reach out for internal infected areas that cannot be treated from the outside; there still are areas in the body that are physically beyond the reach of a doctor. Internal blood clots are one such problem that has been prolonging for a very long time and sometimes can be located in certain areas that are too deep inside, which makes them virtually impossible to reach; deep arteries or veins, narrow internal sections of brain or lungs.*
*In this paper, we propose an approach which puts Nano-robotics into use to deal with the problem of internal blood clots. This method looks at introducing swamps of nanobots in one's bloodstream which acts as carriers of medicines to the sites which are difficult to access for treatment physically. This will not just ease the process but it promises to be far more efficient, precise and less invasive than the currently existing techniques.*

*Keywords— Blood clots, Medical Nanorobotics, Nanomedicine, Thrombolysis.*

## I. INTRODUCTION

Blood clots are gel-like hard mass formed by platelets and fibrin in the body. Blood clots help to prevent excessive bleeding from the body. It basically acts as a mesh and covers the damaged area to ensure the clotting of blood.
It tries to repair a blood vessel, an artery or a vein by acting as a patch till the damage is naturally healed. The blood clot basically reduces the flow of blood out of the damaged region.
Clots are essential in healing of external wounds. They form an outer protective layering over the wound which eventually either dissolves or falls off with time. But sometimes clots form in places inside the body and can cause trouble when the body is unable to naturally dissolve them— like in the arteries that supply the heart or the brain, or in the veins of the legs [1].
Clots inside the body can also be quite mobile, traveling with the bloodstream from place to place. For example, a clot formed in a vein at a place in the body can travel to the lungs with the blow of blood, and one arising in the heart can end up in the brain. Consequently, they can end up in areas where they are too big to travel further and hence obstructs the blood flow. Such silent migrations can have deadly consequences.
Traveling clots are dangerous and can lead to stroke, paralysis or organ failures.
There are two kinds of internal blood clots:
1. Deep Vein Thrombosis: These are the blood clots that are formed in a deep vein usually in the leg or the arm. These are usually fixed in their place.
2. Pulmonary Embolism: If there is a blood clot or thrombus in a deep vein of the leg or arm, it has the potential to break off (embolize) and flow through the veins toward the heart, and into the lung where it can become lodged in a pulmonary artery, which prevents the lung from functioning. Pulmonary Embolism is a medical emergency and can cause serious illness or death [2].

Blood clots affects people from all walks of life. On an average, 274 people die due to blood clots everyday in the USA. This number is larger than people dying from vehicle crashes, AIDS and breast cancer [3].
Due to this, costs due to blood clot healthcare exceed 5 billion dollars per year which hits the economy hard and this cost need to be efficiently tackled.

## II. WHY NANOBOTS FOR BLOOD CLOTS

Blood clots can be fatal at times. The blood streams should be clear of any kind of unwanted clots as they have severe consequences if they get stuck in any artery or vein. To accomplish this task with precision, the small size of a nanobot comes to advantage. As they are small enough to propagate through our bloodstream with ease, they can act directly at the target site. The traditional methods of treating a blood clot, including the use of blood thinners and injecting other clot breaking medicines, can have side effects [4]. This treatment via syringes can become inefficient when the medicine gets diluted while traveling through the blood stream till it reaches its target site and hence a relatively little amount of medicine is delivered. Whereas a nanobot can travel to the exact target site and despite its small capacity to carry a payload, can turn out to be very effective. Particle size has effect on serum lifetime and pattern of deposition. This allows drugs of nanosize to be used in lower concentration and to have an earlier onset of therapeutic action. It also provides materials for controlled drug delivery by directing carriers to a specific location [5]. This also reduces the side effects of the medication.

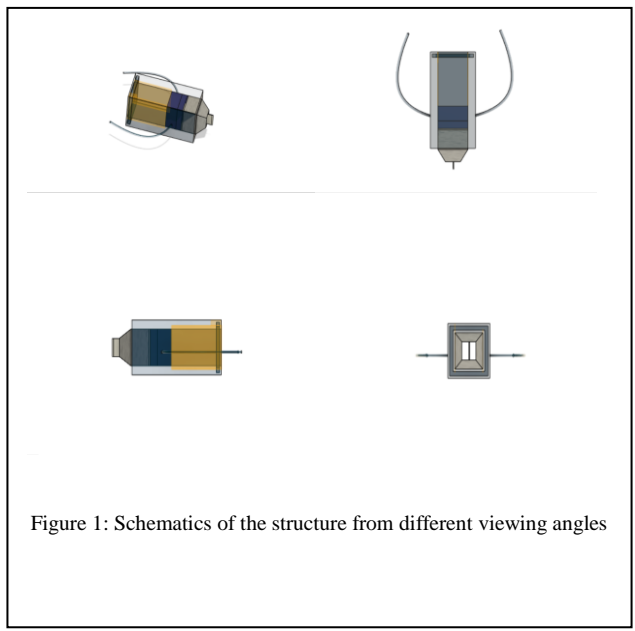

Figure 1: Schematics of the structure from different viewing angles

### III. STRUCTURE

The architecture of nanorobot is similar to that of conventional robots; which consists of driving, frame, sensing and control components.
Driving component provides the force needed in task operation; Frame component and connection component compose the main body of nanorobot; Chemical sensors constitute the sensing component, feedback information and status [6].
The nanobot has a size of 0.1- 10 microns [8] and they will not embolize small blood vessels because the minimum viable human capillary that allows passage of intact erythrocytes and white cells is 3–4 micron meter in diameter, which is larger than the proposed structure.
The structure of the nanorobot proposed by us inspires from the structure of a pharmacyte. Pharmacyte is a self-powered, computer controlled medical nanorobot system capable of digitally precise transport, timing, and targeted-delivery of pharmaceutical agents to specific cellular and intracellular destinations within the human body [7].

The structure of the nanodevice will have the following components:
- A pair of electrodes (which can function as an internal power source by using the electrolytes in the blood [9] )
- Onboard tanks (will act as reservoirs to hold medicinal payloads)
- A set of sensors and onboard chips (which will track the moments of the bot and its surroundings. These sensors will help to decide when and where to drop the medicine by detecting the target site)
- Along with these components, the nanobot will also have appendages which come into use in holding onto surfaces and spraying of medicine.

Combining the above components, our structure is expected to look as shown in Figure 1. The figure is a depiction of the bot from different viewing angles. The simulations and the bot structure constructions were done using the Autodesk Fusion 360 (version 2.0.7830) software.
Assisting the main structural frame will be a propulsion system for propulsion through the blood stream.
When looking at the actual construction of a nanorobot, there are two main directions/ways to it: One is "bottom up", which means building components from atoms or molecule, then assemble components to form machines; the other one is "top down" which means making the nanorobot as a whole and then crafting it with modifications, these will be a novel methods without the limitation of current technology [6].
Both the above ways can be carried out either manually by using special molecular machines for nano structures for synthesis and then sequentially assembling them or by self-assembly process which uses the natural forces to automatically for the structure.
The field of Nanorobotics and its manufacture is currently an emerging field and new breakthroughs are expected in them in the near future. But as of now, a solid production line or methods have not been established to manufacture such custom nano sized robots. Hence, further research and development is required towards the construction side of this approach.

## IV. WORKING

The locomotion of the nanorobot can either be Random or Sensor Signal Guided.
In the random approach nanorobots move passively with the blood flow reaching the target only if they bump into it by obstruction of flow.

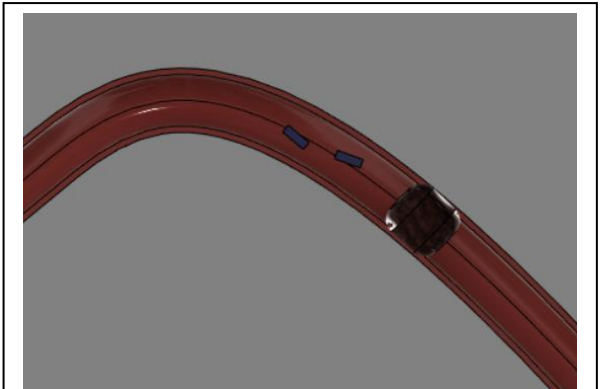

Figure 2 : Nanobots heading towards the target site.

The Sensor Guided approach will make use of the sensor Signals. There is a unit of sensors, each having its specific purpose, which work collectively to aid the movement of the bot in the bloodstream and dominate the process of decision making by providing inputs from the environment. Nanosensors include long distance sensor and short distance sensor, the former navigates the nanorobot to the target tissue, the later locates unwanted cells.

Ultrasonic sensor and light sensor combined with motion strategy can navigate nanorobot in the body [6].

We propose a sensor guided approach here. The movement of the bot can be perfectly synced with the inputs given by Sensors to reach the target site. The thrust will be generated as follows:

The pair of electrodes, placed opposite to each other with blood as the dielectric between them, act as a capacitor. The capacitor will generate magnetic fields which tend to pull conductive fluids (blood in this case) through one end of it which functions as an electromagnetic pump and shoot blood out from the back end. The nanorobot would give an illusion of moving around like a jet airplane [10]. To increase this shooting pressure, the backend of the cavity is made narrow in cross sectional width. A tail will also be a part of this locomotion unit to assist change in directions and to maintain the overall balance of the bot.

The functionality of the Nanorobot comes into play after it successfully reaches the target site. The bot gets mounted to a suitable surface with the use of probes and then the medicine is sprayed over the clot, as sketched in Figure 2.

This medicine can be any of the clot dissolving drug which will be a part of thrombolysis. Drugs like rt-PA: which is currently used for clot dissolving [11] can be useful for this purpose.

Hence, the clot gets dissolved within the bloodstream effectively.

Energy dissipates in the process of driving, operation and information transmission, so sufficient energy supply is required to carry out the full procedure. An internal power sourcing technique will be used to take care of this, making our device self-sufficient to generate power in the bloodstream itself. The electrodes form a battery using the electrolytes found in blood which when completed via a circuit forms a capacitor, which becomes a reservoir for charge.

This way, the complete working of the proposed structure is taken care of inside the bloodstream.

## V. ADVANTAGES

The concept proposed above provides an alternate and an efficient way to combat blood clots. Nanobots are designed to act swiftly in response to combating a blood clot. As soon as it is into the bloodstream, the nanobot swim across the blood to reach the particular destination. The durability of the nanobot still remains the principal advantage of the nanobots. The drug is delivered at the very required site and hence it leaves no room for dilution of the drug through the transportation process. Also, the degree of Invasiveness is eliminated which reduces complexity of the process of treatment and chances of infection.

Therefore, they serve as a complete package for tackling a disease, starting from its diagnosis and regular monitoring to its potential treatment, with the intelligence of a human via its programming and accuracy of a machine [12]. Once the nanobots finishes its required task, it can be recovered and modified by re-programming for making it useful for other tasks as well or just can be easily recycled to perform the same task again. Also, in certain conditions the same nanorobot, once inside the blood stream, can be used to reach multiple target sites in a single operational run.

The system also acts as a self-checker: as it will travel via the bloodstream post dissolving the clot, it counter checks for the blockage being effectively clear of any obstruction.

## VI. DISADVANTAGES

Nanobots are trained on algorithms that work on certain decision-making criteria and are designed for ideal conditions as per the assumptions made by the designers. This may result in malfunctioning of the bots in practical conditions. Obviously Nanobots cannot be as intelligent and situation adaptive as humans and hence are bound to produce some errors in their functioning.

As we are dealing with the potential use of technology inside of living organisms, accuracy is a vital factor to be taken care of as there is little room for errors when it's the question of life and death of the patient. The nanobot must be highly precise since its interference with any cell which is not demanded, can cause side effects/counter response and it should be made sure that the normal functioning of the body is not disturbed. A nanobot being of cellular size can be misinterpreted as a foreign specie if it somehow alters the density of blood or starts reacting with the cells of the body and can trigger a response from the immune system.

It is difficult to design structures with such fine specifications with no margin of error.

Nanobots are after all machines which run on codes and they must be programmed correctly with adequate encryption and security. Otherwise it leaves a window to hackers and terrorists to easily misuse them and make them serve as a potential bioweapon.

Nanorobotics is still an upcoming technology and its efficiency is not yet tested in practical scenarios and with human trials. So the limitations of nanorobots can't really be predicted at this moment as they are virtually unprecedented. Also designing, planning and synthesis of nanotechnology is expensive and requires largely precise equipment [12].

## VII. CONCLUSION

The paper presented a preliminary research on the potential use of nanorobots in getting rid of Blood Clots inside the human body. With the rapid growth in technology, we are looking at a world where robots of a size small enough to travel through the bloodstream will be part of the treatment process for a number of ailments. This way it will present a new dimension to the medical sector and will reduce external human intervention and invasion.

This project can further be extended towards laying the criteria and building algorithms for the identification of the blood clot, once in the bloodstream and also devising path planning algorithms for the nanobots propagation making them reach the target sites efficiently.